\documentclass[journal]{IEEEtran}
\usepackage{cite}

\ifCLASSINFOpdf
\usepackage[pdftex]{graphicx}
\else
\fi
\usepackage[cmex10]{amsmath}
\usepackage{array}

\usepackage[caption=false,font=footnotesize]{subfig}
\usepackage{booktabs}
\usepackage{multirow}
\usepackage{rotating}
\usepackage{color}

\begin{document}
%
%


\title{A Cryogenic Ultra-Low-Noise MMIC-based LNA with a discrete First Stage Transistor Suitable for Radio Astronomy Applications}
%
%
%

\author{Mark~A.~McCulloch,
        Simon~J.~Melhuish,
        and~Lucio~Piccirillo
\thanks{M. A. McCulloch, S. J. Melhuish and L. Piccirillo; - Jodrell bank Centre for Astrophysics, The University of Manchester, Manchester, M13 9PL UK e-mail: (see http://www.jb.man.ac.uk/contact/people.html).}
\thanks{Manuscript received October 11, 2013.}}

\maketitle


\begin{abstract}
In this paper a new design of MMIC based LNA is outlined. This design uses a discrete 100-nm InP HEMT placed in front of an existing InP MMIC LNA to lower the overall noise temperature of the LNA. This new approach known as the Transistor in front of MMIC (T+MMIC) LNA, possesses a gain in excess of 40\,dB and an average noise temperature of 9.4\,K compared to 14.5\,K for the equivalent MMIC-only LNA measured across a 27--33\,GHz bandwidth at a physical temperature of 8\,K. A simple ADS model offering further insights into the operation of the LNA is also presented and a potential radio astronomy application is discussed.
\end{abstract}


%
\IEEEpeerreviewmaketitle

%

\section{Introduction}
%
%
%
%

\IEEEPARstart{L}{OW} noise amplifiers (LNAs) form the most important part of the sensitive coherent receivers that are used in radio astronomy and the advancement that has taken place in our understanding of the universe over the last few decades is in no small part down to the development of ever lower noise LNAs. A good LNA should possess a reasonable amount of gain, which is used to suppress the noise contribution of the following components, and simultaneously it should contribute as little additional noise as possible to the overall system. 

Very low noise Ka-band LNAs were developed by the Jodrell Bank Observatory (JBO) in the previous decade for the European Space Agency's {\it Planck} Low Frequency Instrument (LFI) \cite{davis2009}. These LNAs utilized the common microwave integrated circuit (MIC or hybrid) approach and were based on four discrete Indium Phosphide (InP) 100-nm gate length high electron mobility transistors (HEMTs), with the lowest-noise amplifiers possessing an absolute minimum noise temperature of 5\,K and a band average (27--33\,GHz) of 8.1\,K. 

However, the last three decades have also seen the development of an alternative approach to low-noise amplification: the monolithic microwave integrated circuit (MMIC) LNA. These amplifiers integrate all of the LNA's components (transistors, transmission lines, bias networks) onto a substrate with a large dielectric constant, for example Gallium Arsenide $\epsilon _r \approx 12.9$ or Indium Phosphide (InP) $\epsilon _r \approx 13.1$. The current lowest-noise Ka-band MMICs \cite{tang2006} achieve an average noise temperature of 15.5\,K. 

The integration of these components gives the MMIC LNA several advantages over comparable MIC designs. Their integrated nature reduces the number of components and bond wires required, making them easier and less time-consuming to integrate into a module (chassis). The use of substrates with a large dielectric constant results in substantially smaller LNAs (the MMIC chips are typically $<$ 3\,mm x 2\,mm, whereas a MIC design may measure over 5-cm in length) making them more suitable for multi-pixel arrays. Mass production also increases the probability of fabricating a large number of LNAs with similar characteristics (gain, noise temperature, phase match) offering the potential for far greater control of systematic effects. These characteristics make MMIC-based LNAs an obvious choice for current and future telescope arrays \cite{martin2012, bhaumik2011} where the number of required high-quality low-noise LNAs could measure in the 1000s  \cite{quiet2010}.

However, their integrated nature comes at a price; although individual LNAs are cheap, there are considerable costs involved in the development of high-quality low-noise LNAs, with many wafer runs often being required to achieve the required design specifications. There is also a noise penalty as it is not possible to select the lowest-noise transistors for the first amplification stage. Manufacturing restrictions on the lengths and widths of the transmission lines limit the ability to match the transistors for low noise and the typically used substrates possess a relatively high loss tangent. Once fabricated there is also no opportunity to test and further optimize the design. This latter penalty is particularly true for the first stage input matching network and the bond wires that ground the HEMT's source contact, since small adjustments to the dimensions of these components can have a large impact on the noise and gain performance.

The purpose of this paper is to follow on from our first report of the successful hybridization of MIC and MMIC technologies \cite{mcculloch2012} with a more detailed look at the amplifier and to report the improved performance that has resulted from additional cooling to a physical temperature of 8\,K. A model of the amplifier has also been developed and is used for discussion of the current design flaws, the underlying physics and what the next stage in the LNA's development should be. We were originally motivated to develop this LNA in an effort to offer a relatively easy way of reducing the noise temperature of MMIC-based LNAs. This approach will be of benefit to radio astronomy in general, since it should allow for a greater use of cheaper commercial MMICs. It could also prove particularly useful for future CMB experiments where arrays with a large number of pixels are planned and one of the priorities will be to detect the very weak B-mode polarization signal. 


\section{Design Basis}

\subsection{Cascaded Systems}
The noise temperature of a cascaded system is given by the familiar Friss equation (\ref{eqn_cascaded_noise}) \cite{friss1944}, which shows that the overall noise temperature of such a system is dominated by both the noise temperature $(T_1)$ and the gain $(G_1)$ of the first component.

\begin{equation}
 \label{eqn_cascaded_noise}
 T_n = T_1 + \frac{T_2}{G_1} + \frac{T_3}{G_{1}G_{2}} + \dots
\end{equation}


Thus placing a discrete transistor in front of a MMIC should lead to two beneficial effects. Firstly, the MMIC's contribution to the noise temperature of the overall amplifier will be suppressed by the gain of this transistor. Secondly, the LNA's noise temperature will be determined primarily by the noise temperature of the transistor. Therefore if the transistor possesses a noise temperature that is less than the noise temperature of the MMIC, the resulting LNA will possess an overall noise temperature only slightly higher than that of the transistor.

\subsection{Linear Two-Ports}

The advantages of a discrete matching network and discrete transistor can be 
illustrated  by considering the work of Pospieszalski \cite{pospieszalski1986}, 
who showed that the minimum noise $(T_{min})$ of any linear two-port device is given by (\ref{eqn_minimum_noise}), 
where $T_0$ is the standard temperature (290\,K), $Z_s$ is the source impedance, 
$Z_{opt}$ is the optimum source impedance, $R_s$  is the source resistance and 
$R_{opt}$ is the optimum source resistance. $N$ is given by (\ref{eqn_n}), 
where $G_{opt}$ is the optimal conductance and $R_n$ and $G_n$ are the noise resistance and conductance respectively.

\begin{equation}
 \label{eqn_minimum_noise}
 T_n = T_{min} + NT_0\frac{|Z_s - Z_{opt}|^2}{R_sR_{opt}}
\end{equation}

\begin{equation}
 \label{eqn_n}
 N = G_nR_{opt} =R_nG_{opt}
\end{equation}

(\ref{eqn_minimum_noise}) shows that it should be possible to match all devices 
to a given input impedance for which $T_n$ will equal $T_{min}$. The transistor's matching network can be designed to do this. 
However, whilst this is relatively ``easy'' to do with discrete devices, 
the integrated nature of MMICs
means that it is not usually possible to match their 50\,$\Omega$ inputs ideally to the 
input impedance of the first transistor. 
Indeed it has been shown that even at lower frequencies the on-chip matching 
network contributes several degrees more to the overall noise temperature 
of the amplifier than an off-chip network \cite{aja2011}. 
These issues become even more apparent when you consider the use of LNAs at cryogenic temperatures. 
Typically an LNA designer may need to adjust matching network parameters 
iteratively to converge on a design optimized for cryogenic operation.
For MIC LNAs this is much easier to achieve than for MMICs, 
for which the design is fixed at the time of the wafer run,
although variations on a design may be accommodated on a wafer.

Consequently we developed a design for an LNA known as the Transistor in front of MMIC (T+MMIC) that exploits not only an off-chip matching network, 
but also utilizes a discrete transistor for the first stage of amplification. 
This facilitates:
\begin{itemize}
 \item An accurate match to the first stage transistor, optimizing noise performance.
 \item Ease of modification to the design for cryogenic use.
 \item Suppression of the noise of the less-than-ideally matched MMIC.
\end{itemize}


\section{The Design}
\subsection{The Active Devices}

\begin{figure}[!h]
  \centering
  \subfloat[]{\includegraphics[width=1.13in]{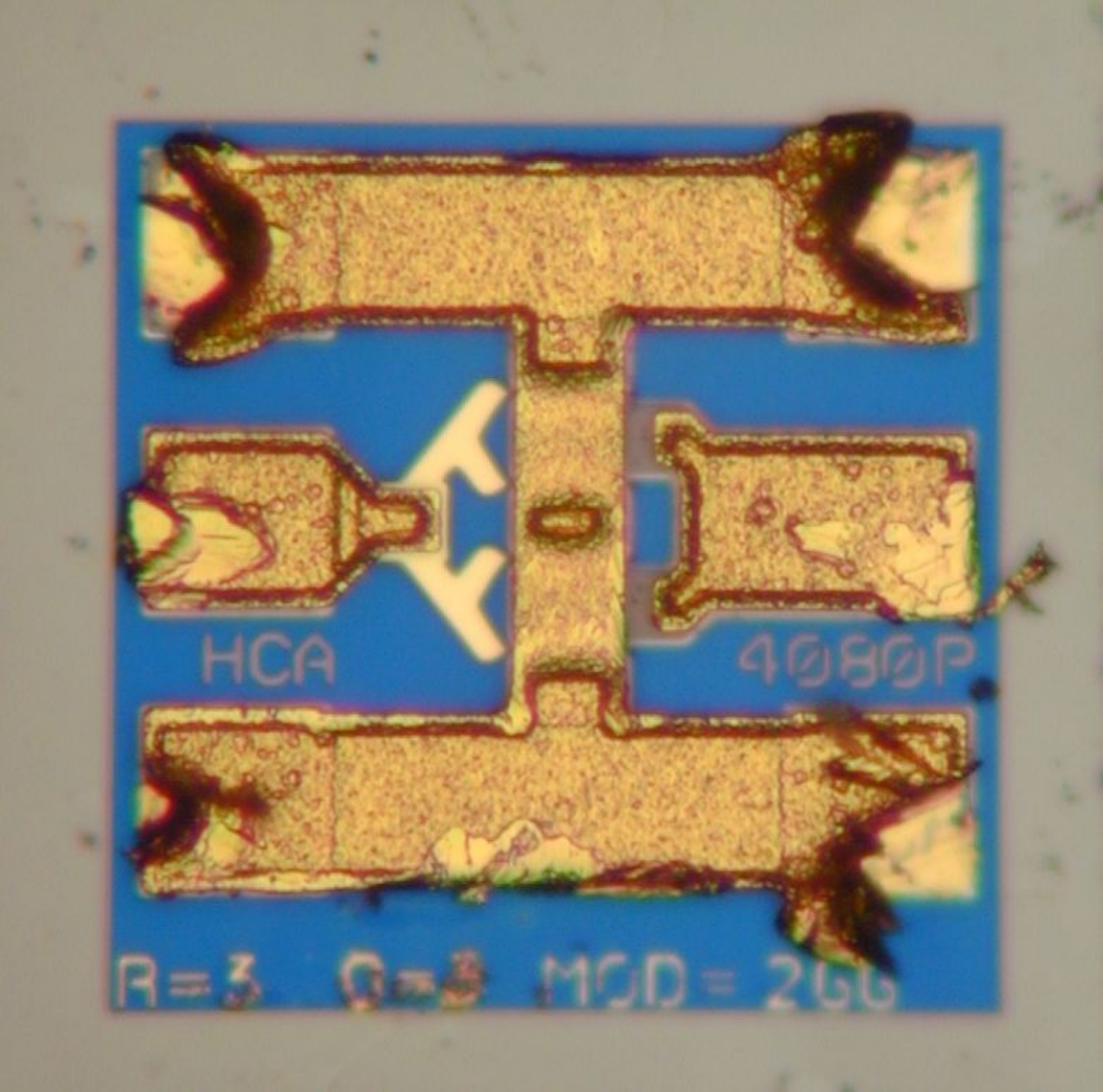}\label{fig_cryo3}}             
  \hspace{1mm} \subfloat[]{\label{fig_mmic}\includegraphics[width=1.5in]{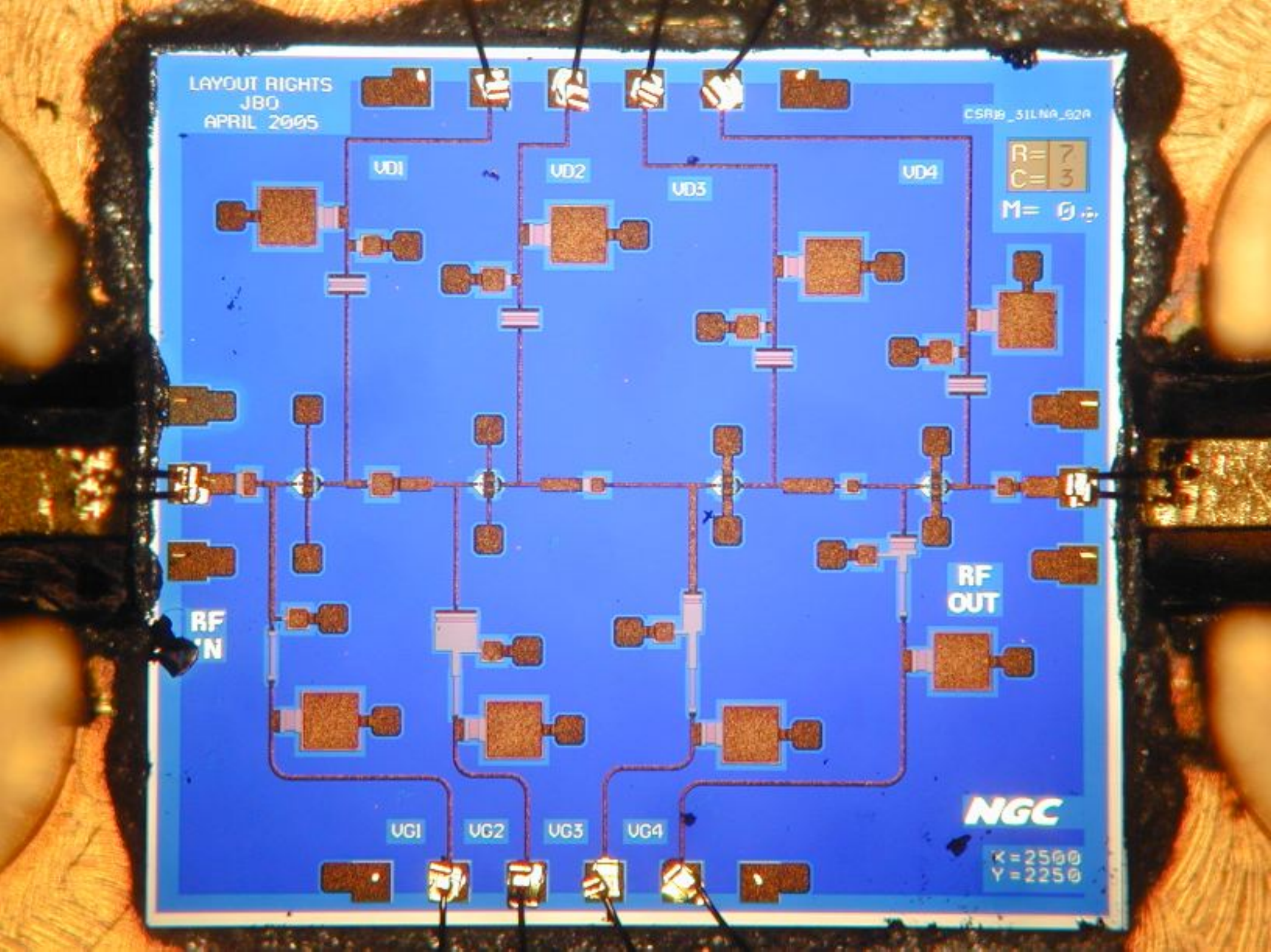}}
  \caption{The Active Devices: (a) a Cryo-3 transistor with approximate dimensions $350\, \mu m \hspace{1pt} \text{x} \hspace{1pt} 350\, \mu m $, (b) a FARADAY MMIC LNA with dimensions $2500\, \mu m \hspace{1pt} \text{x} \hspace{1pt} 2250\, \mu m$.}
  \label{fig_active_devices}
\end{figure}

The T+MMIC LNA is based around two active components: a high-quality transistor and a good MMIC LNA. 
The transistor (Fig. \ref{fig_cryo3}) is a $4\times20\,\mu$m, 100-nm gate-length InP HEMT, that had originally been supplied to JBO by JPL Pasadena 
for use in the {\it Planck} project. The transistor was fabricated under the Cryogenic HEMT Optimization Program (CHOP) \cite{shell2007}, and originated from wafer run 3. 
These particular transistors, known as Cryo-3, still offer state-of-the-art noise performance and the use of such a transistor allows the T+MMIC LNA 
to be compared with the {\it Planck} LFI Ka-band LNAs.


The MMIC LNA (Fig. \ref{fig_mmic}) was developed as part of the European Commission's FARADAY project \cite{kettle2005}.
Since this was a radio astronomy project the MMIC LNA already possessed a good cryogenic noise temperature, 
typically around 20\,K, and a gain in excess of 40\,dB across its 26--36\,GHz operating band. 
The FARADAY MMICs were fabricated on InP by Northrop Grumman Space Technologies (NGST).
They utilize four $4\times30\,\mu$m gate width, 100-nm gate length transistors.

\subsection{Module Design}

\begin{figure}[!h]
 \centering
 \includegraphics[width=3.5in]{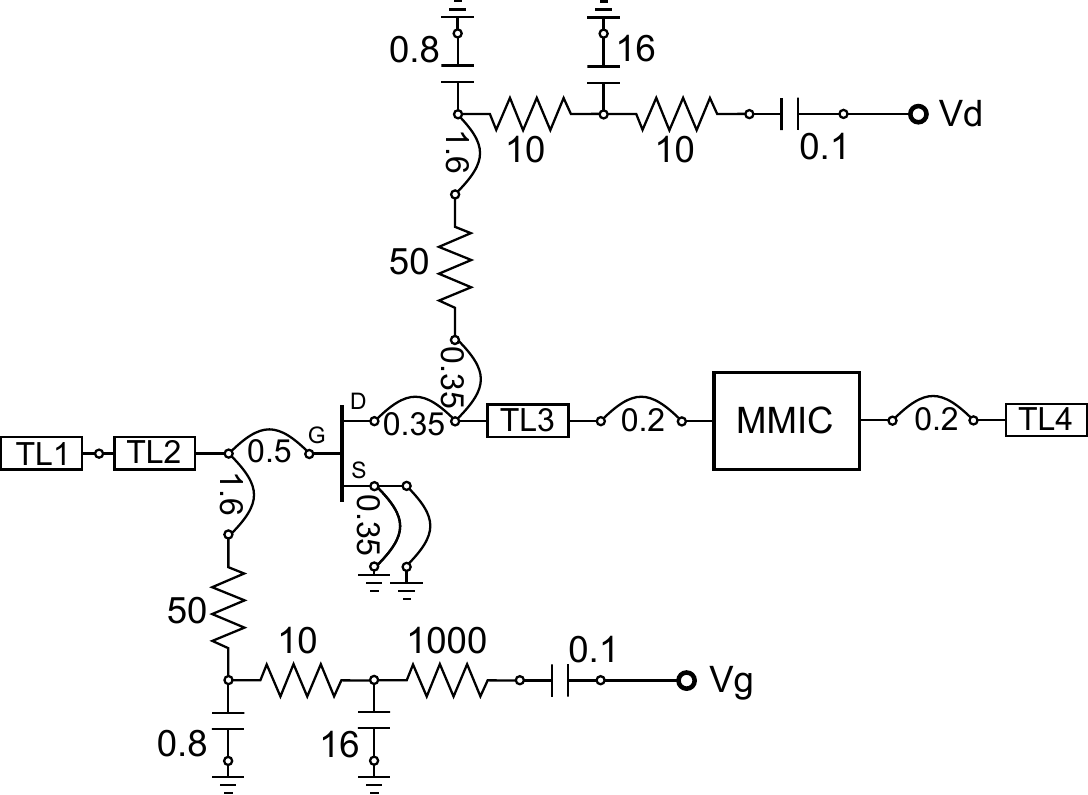}
 \caption{RF circuit schematic for the T+MMIC LNA. Resistors and capacitors have units of $\Omega$ and pF respectively, bond wire lengths are in mm. Transmission lines (TL) 1, 3 \& 4 all have an impedance of 50\,$\Omega$ and lengths of 1.05, 7.2 and 31\,mm respectively. TL\,2 which is used to match the the source impedance to the transistor's input impedance for minimum noise temperature  has an impedance of 23\,$\Omega$ and a length of 0.85\,mm.}
 \label{fig_layout}
\end{figure}

\begin{figure}[!h]
 \centering
 \includegraphics[width=3.5in]{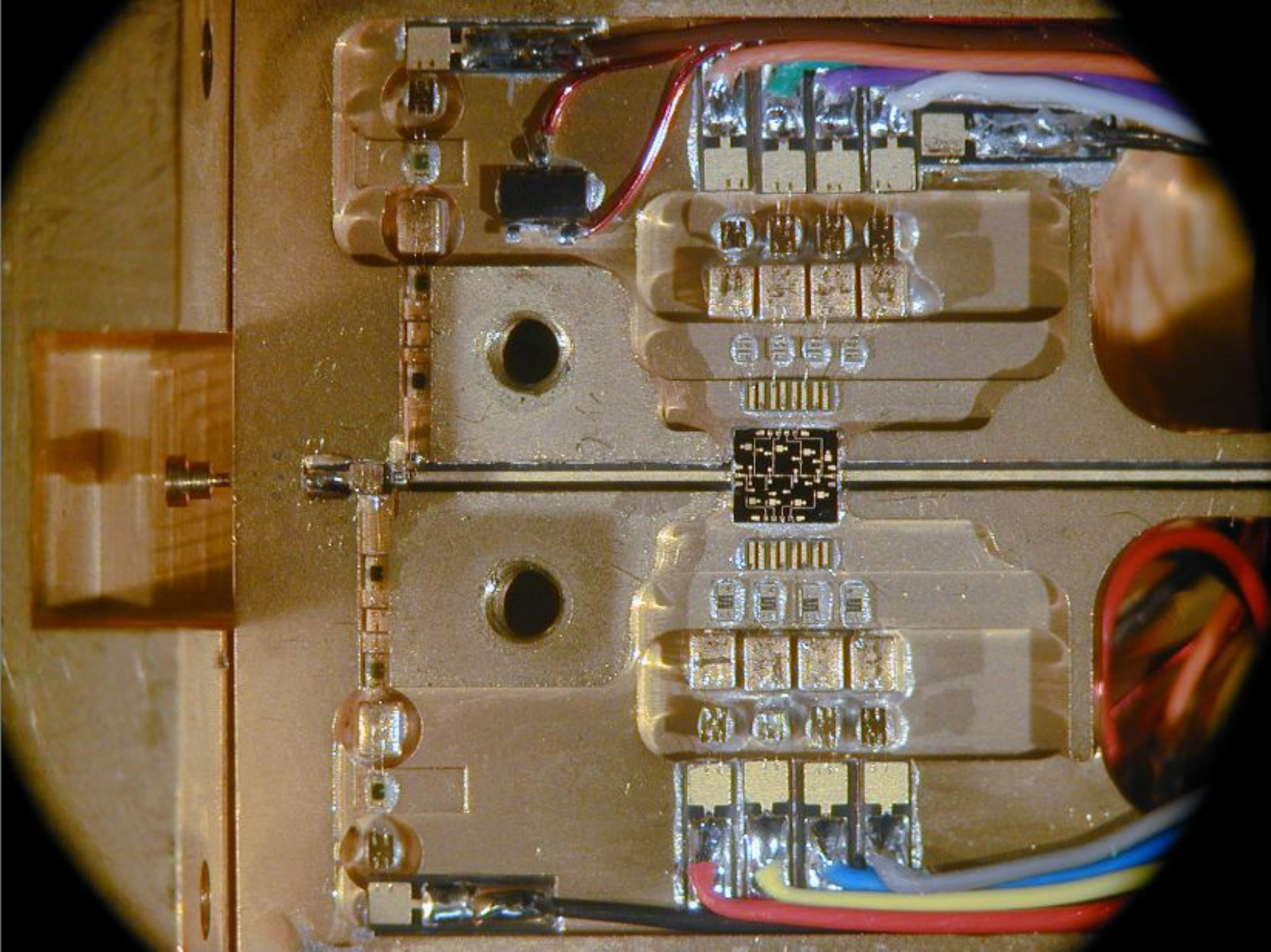}
 \caption{The assembled LNA. From left to right: probe, input matching network, gate bias, transistor, drain bias, 50\,$\Omega$ microstrip line, MMIC, 50\,$\Omega$ output transmission line. A schematic of the RF circuit can be seen in Fig. \ref{fig_layout}}
 \label{fig_assembly}
 \end{figure}

For ease of assembly and to allow the use of pre-existing designs and components the module that houses the T+MMIC LNA  is a merger of two existing JBO-designed LNA modules. 
The transistor section is based around the first stage of the {\it Planck} 30-GHz front end LFI LNA
and the MMIC section is based on a Faraday MMIC test module that was developed as part of the QU-Instrument-JOint-Tenerife Experiment (QUIJOTE) \cite{rubino2010}. 
This approach enabled the use of the existing {\it Planck} Cryo-3 input matching network, 
which avoided the need for a re-design of the module, though this did restrict the bandwidth of the amplifier to 27-33\,GHz. 
The merger of two modules also resulted in the transistor and the MMIC being connected by a rather long ($\sim$7.2\,mm) transmission line.
This compromise was necessary in order to ensure adequate space in the module for the transistor bias circuitry. The module was machined in-house from brass and was gold plated.
The internal components are connected to the waveguide ports by broadband microstrip to waveguide probe transitions. 
The layout of the prototype RF circuit can be seen in Fig. \ref{fig_layout}.
The assembled module can be seen in Fig. \ref{fig_assembly}. 
The gold-plated microstrip lines are fabricated on a 76-$\mu$m Polyflon Cuflon substrate \cite{cuflon}, which has an electrical permittivity and a dielectric loss tangent of 2.05 and 0.00045 respectively.

\subsection{Theroretical Performance}

Since the {\it Planck} amplifier used a Cryo-3 transistor for its first 2 stages we can estimate the average noise temperature of our LNA by using (\ref{eqn_cascaded_noise}) and the average noise performance of both the {\it Planck} amplifier and the MMIC. Considering the first two stages (from (\ref{eqn_cascaded_noise}) the third and forth stages are negligible) of the {\it Planck} amplifier the noise temperature can be written as the following:

\begin{equation}
 \label{eqn_planck}
 T_{Planck} = x + \frac{x}{G}
\end{equation}

Where $T_{Planck}$ is the average noise temperature of the {\it Planck} amplifier (8.1\,K), $x$ and $G$ are the noise temperature and gain (8\,dB) of the Cryo-3 respectively. For the T+MMIC amplifier the following can also be written:

\begin{equation}
 \label{eqn_lna}
 T_{LNA} = x + \frac{T_{MMIC}}{G}
\end{equation}
 
Where $T_{MMIC}$ is the average noise temperature of the Faraday MMIC ($\approx$20\,K). Re-arranging (\ref{eqn_planck}) and (\ref{eqn_lna}) and eliminating $x$ gives the expected noise temperature of the T+MMIC LNA as $\approx$10K (\ref{eqn_expected_noise}).

\begin{equation}
 \label{eqn_expected_noise}
 T_{LNA} = \frac{T_{Planck}}{1 + \frac{1}{G}} + \frac{T_{MMIC}}{G} \approx 10\text{K}
\end{equation}


\section{Noise Test Equipment}

\subsection{RF System}

The noise test set-up has been previously used to develop LNAs for the {\it Planck} Satellite and the Merlin array of radio telescopes. It consists of an HP (now Agilent Technologies) 8350B sweep generator, with an 83550A plug-in module and a 83554A mm wave source module to provide an LO to a down converting mixer, and an 8970B noise figure meter on the IF. For room-temperature measurements an Agilent R347B noise source is used to supply two different levels of input noise to the LNA. Cryogenic measurements were made using the ``hot" and ``cold" load approach where a variable temperature microwave-absorbing (and therefore emitting) waveguide load is coupled to the input of the LNA. Since the load is placed within the cryostat and we are able to measure the temperature of the load with great accuracy our results should have a similar level of accuracy. $S_{11}$ measurements of the load show that the match to the LNA is reasonable ($S_{11}<$-15\,dB), thus we are confident that $T_{phys}\simeq T_{RF}$.

\subsection{The Cryostat}

To perform the cryogenic noise measurements the LNA was placed in a cryostat that had previously been developed for sub-2-K physical temperature noise temperature investigations \cite{melhuish2012}. The cryogenic cooling system is built around a Pulse Tube Cooler (PTC), manufactured by Sumitomo Heavy Industries (SHI), with two cooling stages; the first cools a radiation shield to $\sim$50\,K and pre-cools the 2nd stage, whilst the 2nd stage cools (with no LNA) to $\sim$3\,K. Further cooling is available though it is not used for this investigation.

To enable the LNA's physical temperature to be controlled a TO220 resistor is attached to the LNA's body to allow its temperature to be increased. An identical approach is used to control the temperature of the hot / cold load. Both the LNA and the load are attached to the 3-K stage by means of a thermal switch, which allows for a more rapid adjustment in temperature than would be possible with only a resistor and a weak thermal link. Waveguide is used to carry the RF signal out of the cryostat. The waveguide is composed of two sections; a section of thin-walled gold-plated stainless steel waveguide (SS WG) and a section of brass waveguide (BR WG). These are used rather than copper as they offer reduced thermal conduction. The load is connected to the amplifier via a waveguide thermal break and a length of stainless steel waveguide; again this is to isolate thermally the LNA from the load. 

\begin{figure}[!h]
 \centering
 \includegraphics[width=2.5in]{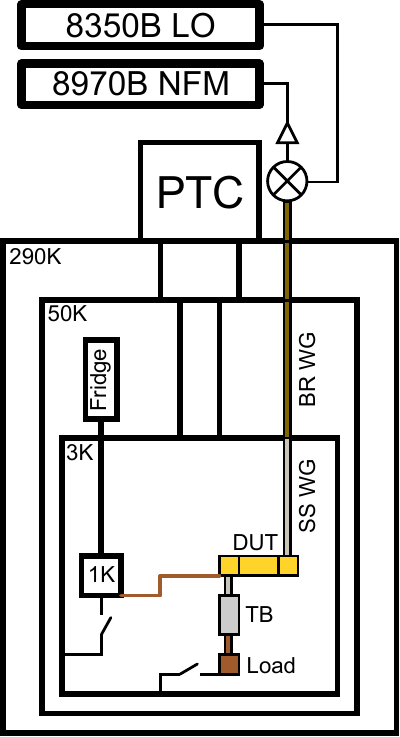}
 \caption{The cryostat's layout when configured for noise temperature measurements. An additional waveguide can be installed for S parameter measurements. The optional 1\,K fridge is also shown. Reproduced from \cite{melhuish2012}.}
 \label{fig:cryostat}
 \end{figure}

A cryogenic readout / control unit that we developed previously
from a device used for the QUAD telescope is used to monitor
key temperatures within the system and to control the various
heaters. A PID control loop sets load and LNA temperatures.
We use silicon diode and ruthenium-oxide thermometers, calibrated
against a rhodium-iron standard and GRTs to monitor the temperature.
Fig. \ref{fig:cryostat} shows the layout of the cryogenic
system for noise temperature measurements.

\section{Performance}

\begin{figure}[!h]
  \centering
  \subfloat[]{\includegraphics[width=3.5in]{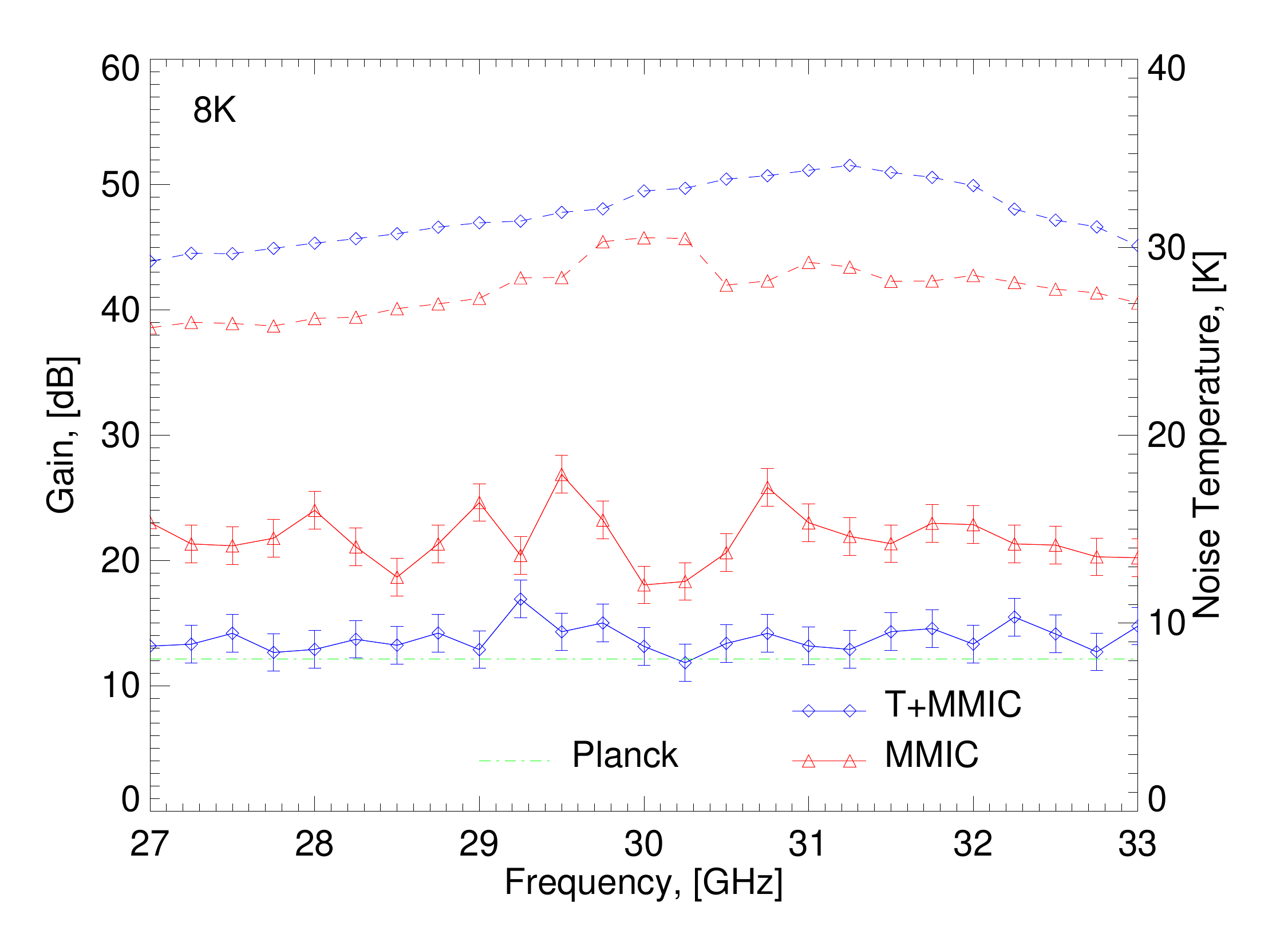}\label{fig_perform_cryo}}  \\
           
  \subfloat[]{\label{fig_perform_rt}\includegraphics[width=3.5in]{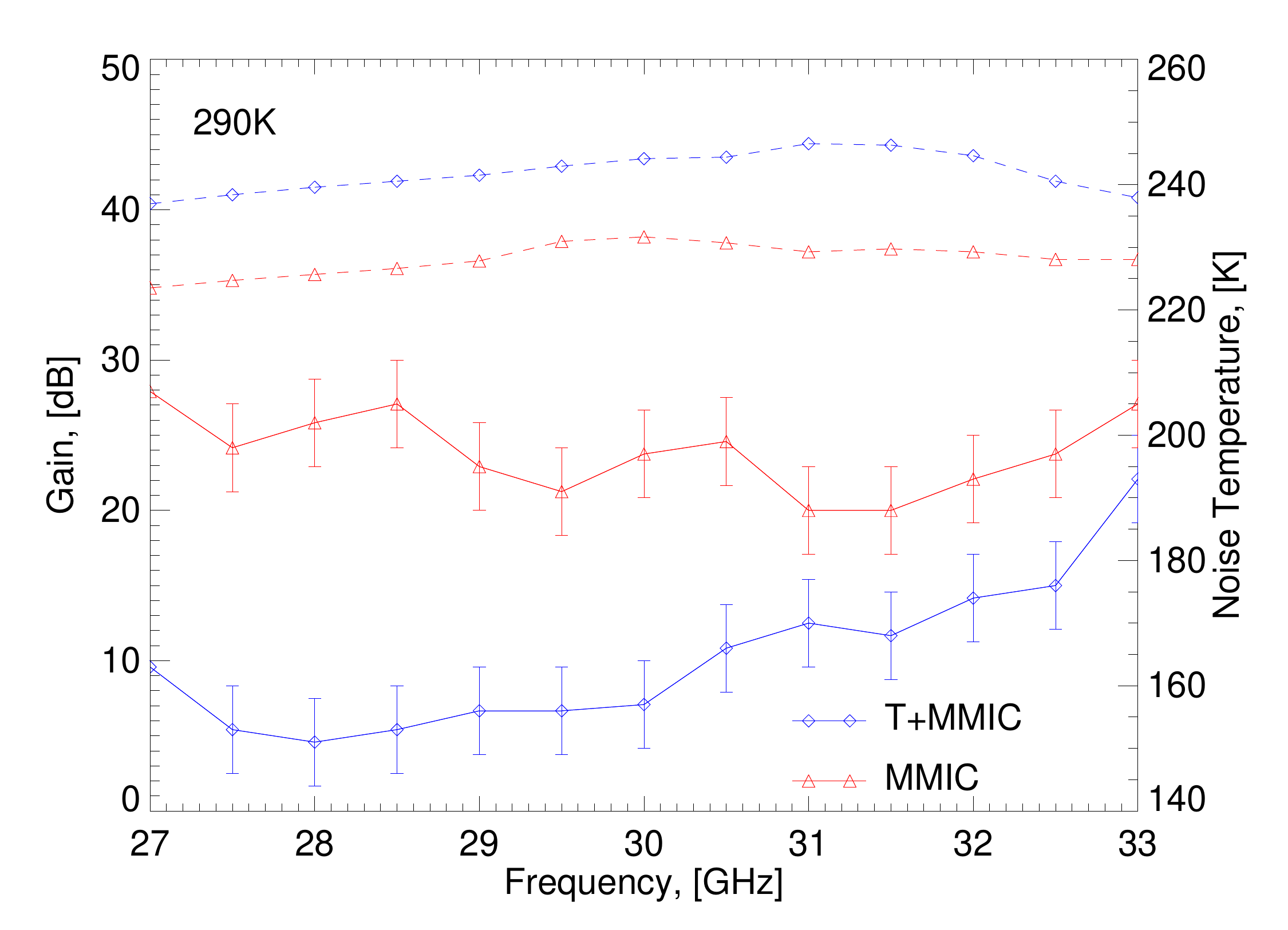}}
  \caption{The T+MMIC LNA's performance at (a) 8\,K  and (b) 290\,K, compared with measurements from an LNA consisting of only an MMIC (same design
and wafer).}
  \label{fig:performance}
\end{figure}

The performance of the T+MMIC LNA was measured at a physical temperature of 8\,K and the results are shown in Fig. \ref{fig_perform_cryo}. For comparison purposes the results for a Faraday only MMIC LNA are also shown. For completeness, the room temperature performance is also shown in Fig. \ref{fig_perform_rt}. In all cases the amplifiers were biased for minimum noise. To highlight the effectiveness of this technique the {\it Planck} LFI average noise temperature (8.1K) is also shown, though it should be noted that this was  measured at a physical temperature of 20\,K. The T+MMIC LNA was measured at 8-K owing to recent results \cite{melhuish2012} where an LNA was cooled to 2-K that showed that noise temperature continues to decrease with decreasing physical temperature. We estimate that at cryogenic temperatures the uncertainty in our noise measurements is $\pm 1$\,K whilst at room temperature we estimate that this increases to $\pm 10$\,K. These estimates are based on repeated observations and are consistent with other measurements that have used similar techniques \cite{munoz1997}.

\section{Modeling}

For further insight into the operation of the T+MMIC LNA a model of the amplifier was produced using Agilent's Advanced Design System (ADS). The transistor stage (transistor, bias networks, input and output transmission lines) is fully simulated in ADS, the outputs of which are then combined with the MMIC's S parameters and noise behavior to give the LNA's overall characteristics. 

\subsection{The Transistor Stage }

\subsubsection{Cryo-3 Transistor}
\begin{figure}[!h]
 \centering
 \includegraphics[width=3.5in]{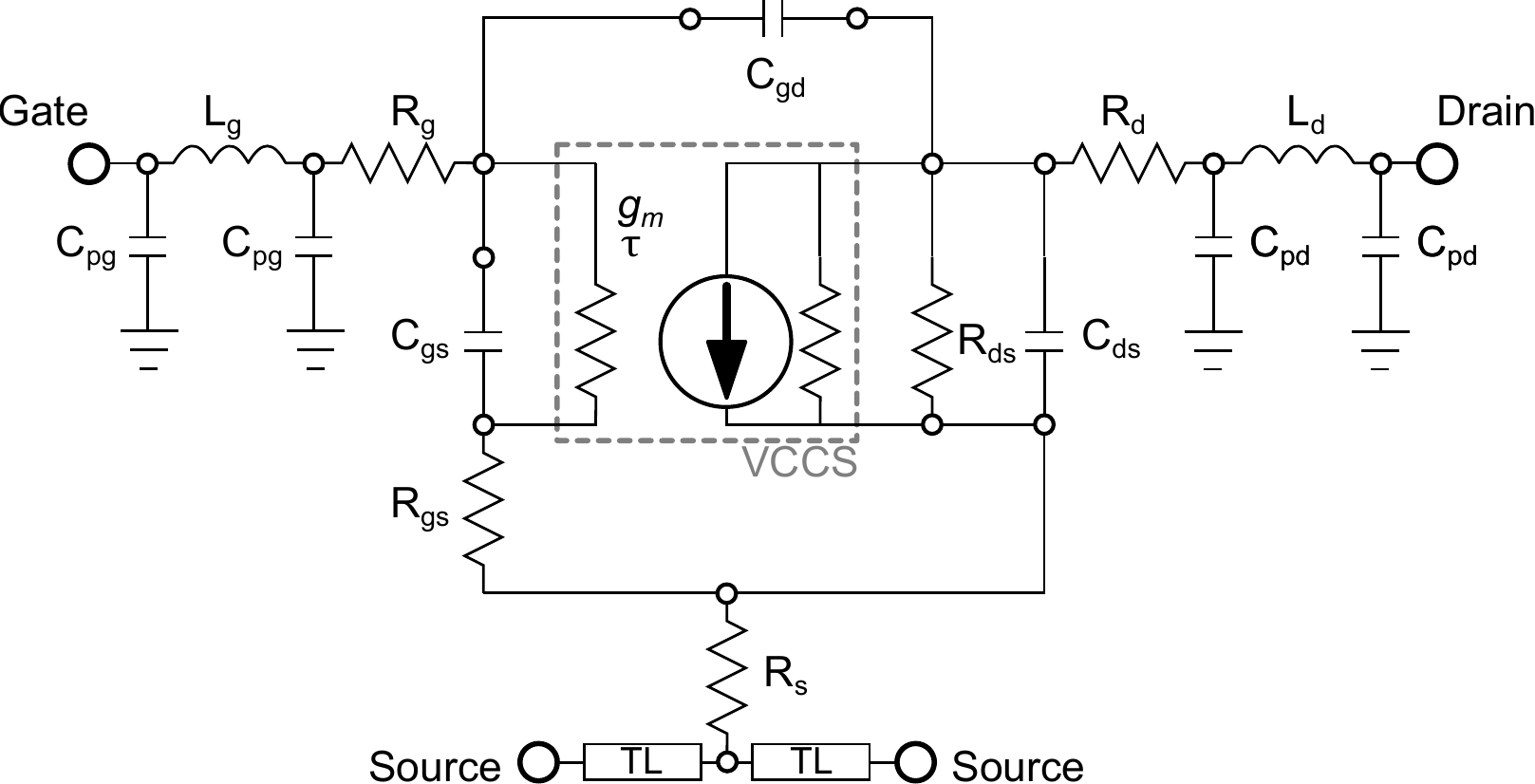}
 \caption{A transistor equivalent circuit, suitable for use in Agilent's Advanced Design System. Developed by M. Pospieszalski \cite{pospieszalski2004}.}
 \label{fig_transistor}
 \end{figure}

The Cryo-3 transistor was simulated using a modified version (Fig. \ref{fig_transistor}) of the standard 15-parameter equivalent circuit model (see \cite{berroth1990, dambrine1988,tayrani1993} for a discussion on parameter extraction). This particular version was developed by M. Pospieszalski \cite{pospieszalski2004}. The equivalent circuit parameters were measured as part of the {\it Planck} project and are shown in Table \ref{tab:cryo3_parameters}. Further details can also be found in \cite{pospieszalski2012}. For the room temperature simulations the passive extrinsic and intrinsic parameters are left unchanged from those of the cryogenic simulation \cite{pospieszalski1993}, whilst the Pospieszalski noise equivalent temperatures \cite{pospieszalski1989} $T_d$, $T_g$ and $T_a$ are adjusted appropriately. For room temperature modeling the transconductance $g_m$ is decreased by $\sim$20\% over the cryogenic temperature value \cite{pospieszalski1997}. The Voltage Controlled Current Source (VCCS) illustrated in Fig. \ref{fig_transistor} is the approach used in ADS to 
model a 3-port transistor. The other parameters are simulated by using ideal lumped components. The two transmission lines denoted TL are used to represent the inductance of the source pads. This inductance is modeled through the use of ideal microstrip lines with impedance = 20\,$\Omega$, length = 63.5\,$\mu$m, $\epsilon_r$= 13.1.

\subsubsection{Bias Networks}
The components in the transistor's bias networks are also simulated using lumped components, whilst the transmission lines are simulated using microstrip components.

\begin{table}[htbp]
  \centering
  \caption{Cryo-3 equivalent circuit parameters}
    \begin{tabular}{crcc}
    \toprule
    \multicolumn{2}{c}{} & \hspace{7mm} 8\,K   & \hspace{7mm} 290\,K \\
    \midrule
    \multirow{2}[3]{*}{Bias}&  Vd    &\hspace{7mm} 0.9\,V  & \hspace{7mm} 1.2\,V \\ 
          &  \hspace{7mm}Ids   & \hspace{7mm} 2\,mA   & \hspace{7mm} 6\,mA \\  \midrule
    \multirow{3}[2]{*}{Noise}  & Ta    & \hspace{7mm} 8\,K   & \hspace{7mm} 297\,K\\
          &  Tg    & \hspace{7mm} 8\,K    & \hspace{7mm} 20\,K \\
          &  Td    & \hspace{7mm} 400\,K    & \hspace{7mm} 1500\,K \\ \midrule
    \multirow{1}[2]{*}{Gain} & $g_m$    & \hspace{7mm} 80\,mS   & \hspace{7mm} 67\,mS\\ \midrule \hline
 & & & \\  
	\end{tabular}
    \begin{tabular}{rcrc}
    \multicolumn{4}{c}{Extrinsic Parameters} \hspace{9mm}  Intrinsic Parameters \hspace{3mm}\\
    \midrule
            Rg    & \hspace{7mm} 1\,$\Omega$  \hspace{7mm} &Cgs & \hspace{6.5mm} 52\,fF \\
            Rd    & \hspace{7mm} 5\,$\Omega$  \hspace{7mm} &Cgd&\hspace{6.5mm}  24\,fF \\
           Rs    & \hspace{7mm} 2.2\,$\Omega$ \hspace{7mm} &Cds & \hspace{6.5mm} 10\,fF \\
           Cpg   & \hspace{7mm} 4.6\,fF \hspace{7mm} &Rds& \hspace{6.5mm} 135\,$\Omega$ \\
           Cpg   & \hspace{7mm} 4.6\,fF \hspace{7mm} &Rgs& \hspace{6.5mm} 4\,$\Omega$ \\
           Cpd    & \hspace{7mm} 12\,fF \hspace{7mm} &$\tau$ & \hspace{6.5mm} 0.6\,psec \\
           Cpd    & \hspace{7mm} 4.6\,fF \hspace{7mm} & & \hspace{6.5mm}  \\
           Lg    & \hspace{7mm} 9\,pH \hspace{7mm} &$$ & \hspace{6.5mm}  \\
           Ld    & \hspace{7mm} 16\,pH \hspace{7mm} & & \hspace{6.5mm}  \\
    \bottomrule
    \end{tabular}%
  \label{tab:cryo3_parameters}%
\end{table}%

\subsubsection{Bond Wires}

\begin{figure}[!h]
  \centering
  \subfloat[]{\includegraphics[width=2.0in]{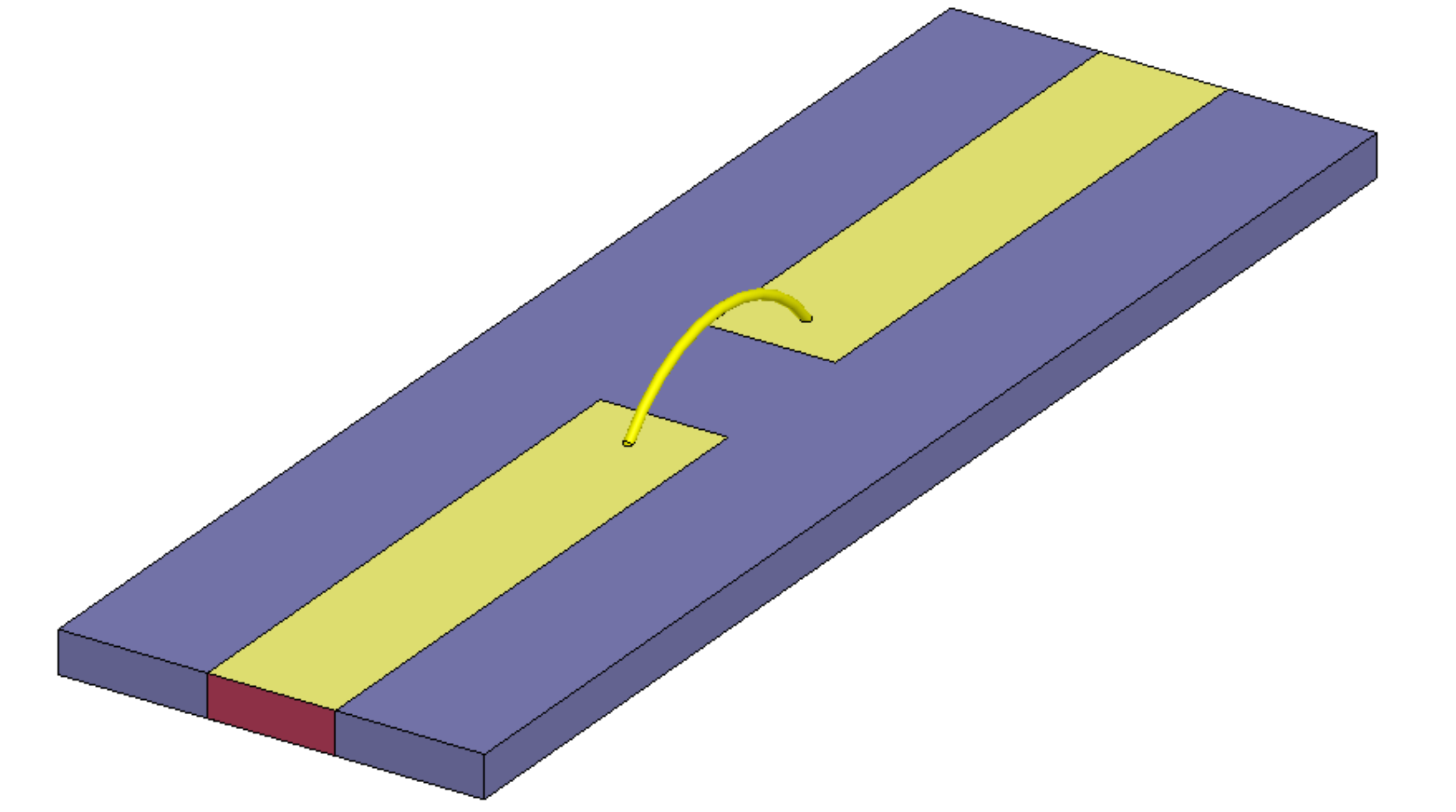}\label{fig:bond-wire}}     
       
  \subfloat[]{\label{fig:bond-wire-smith}\includegraphics[width=3.0in]{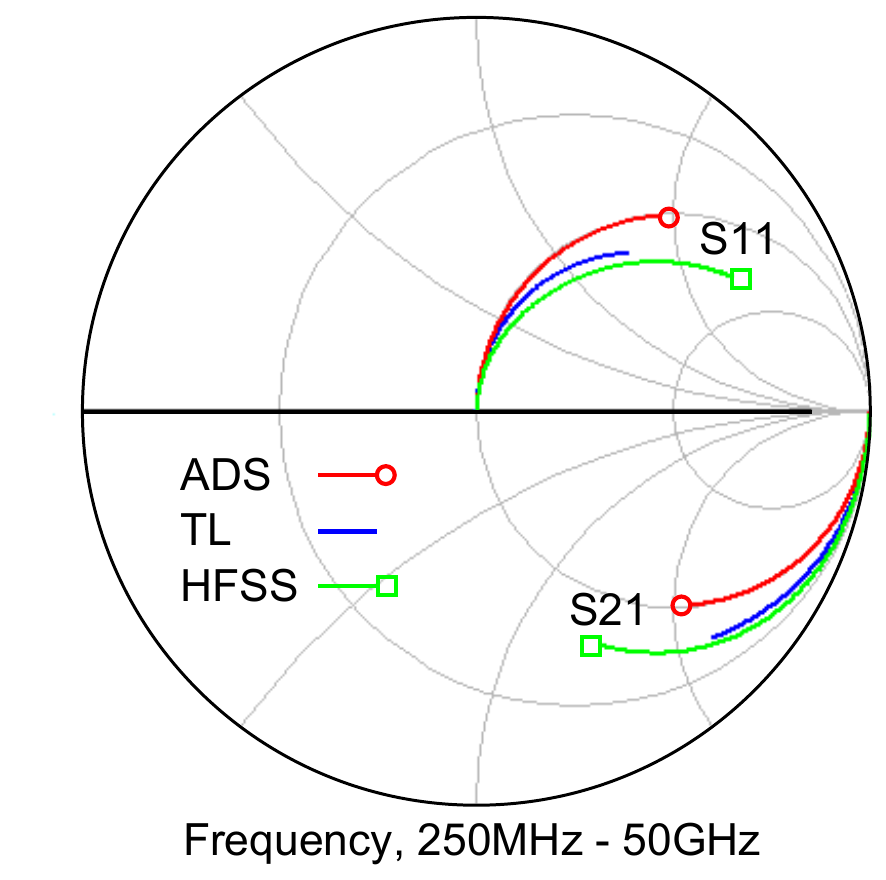}} 

  \caption{(a) An HFSS model of a bond wire, ready for simulation. (b) The results of 3 different (ADS, HFSS and an ideal transmission line) approaches to the modeling of a 500\,$\mu m $ bond wire.}
  \label{fig:bond-wires}
\end{figure}

Although ADS includes bond wire simulation tools, the bond wire model only takes into account the wire's inductance, not the capacitive effects that arise from the surrounding metallic structure and the bonding points themselves. Since these effects can have a significant impact on the performance of the RF circuit, considerable work has been done to investigate and describe the behavior of bond wires \cite{lee1995,alimenti2001} at mm-wave frequencies. The quasi-static model outlined in \cite{alimenti2001} utilizes 4 transmission line elements to help model the wire's behavior. Ansys' High Frequency Structure Simulator (HFSS), an EM simulator, can also be used to simulate the behavior of bond wires. However, like the quasi-static model the wire's profile needs to be well known, which is not always possible, as was the case here. Therefore a simpler single transmission line approach, where the bond wires are simulated using high impedance ($150 \Omega$), ideal ($\epsilon_r$ = 1) transmission lines was used. This approach had proved useful during the development of the {\it Planck} LNAs \cite{pospieszalski2004}. Fig. \ref{fig:bond-wires} illustrates the difference between a 500\,$\mu$m (linear length) bond wire that was simulated in ADS using the bond wire 
simulation component, an equivalent bond wire (Fig. \ref{fig:bond-wire}) simulated in HFSS and an ideal 150\,$\Omega$ transmission line. Fig. \ref{fig:bond-wire-smith} shows that a single high impedance transmission line still represents a good approximation to the HFSS simulation. 

\subsection{Faraday MMIC S Parameters}

\begin{figure}[!h]
 \centering
 \includegraphics[width=3.5in]{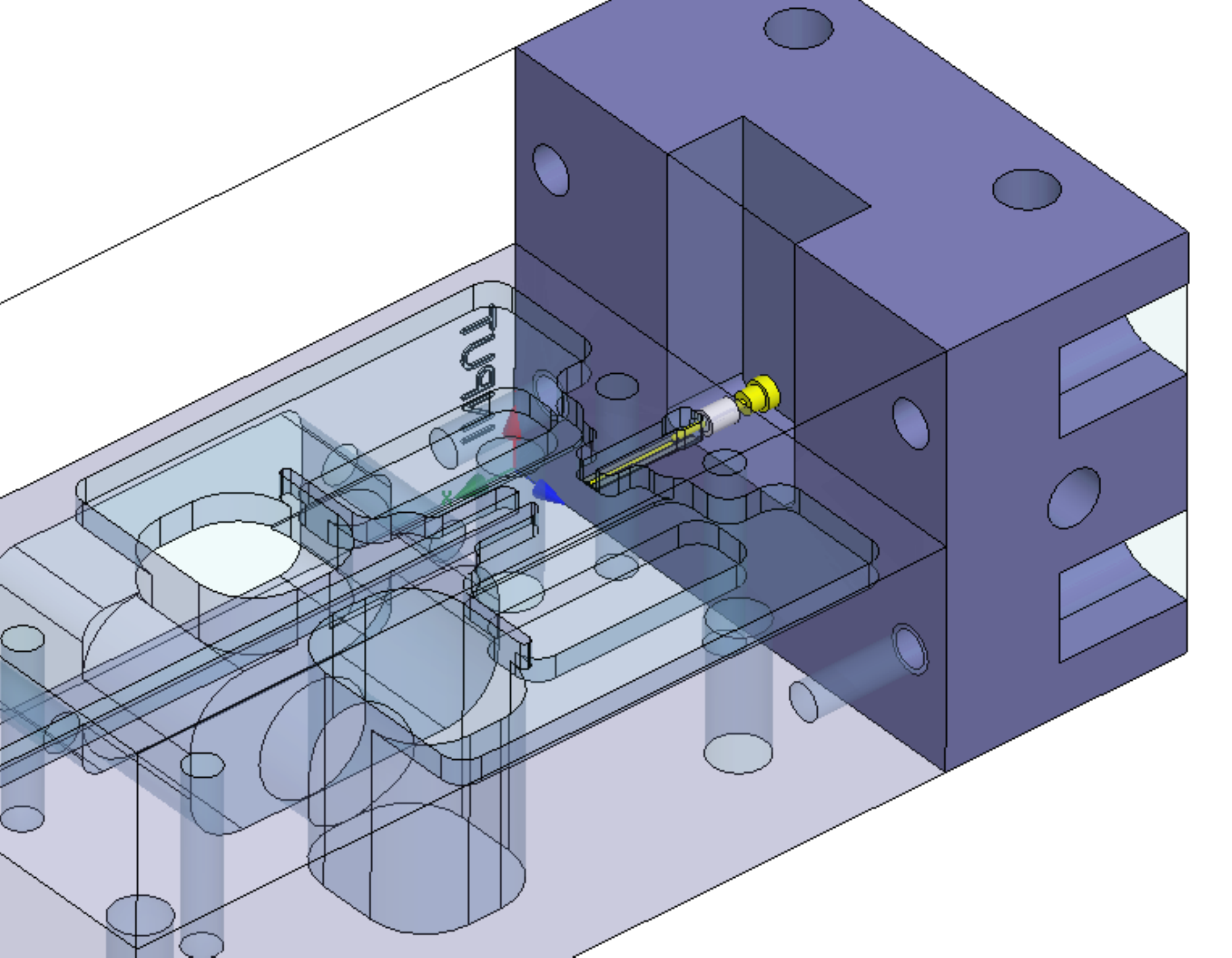}
 \caption{The HFSS model of the input broadband waveguide to microstrip transition. In the case of the cryogenic measurements the input and output stainless steel and brass waveguides are also simulated in HFSS.}
 \label{fig:probe}
 \end{figure}

\begin{figure}[!h]
 \centering
 \includegraphics[width=3.5in]{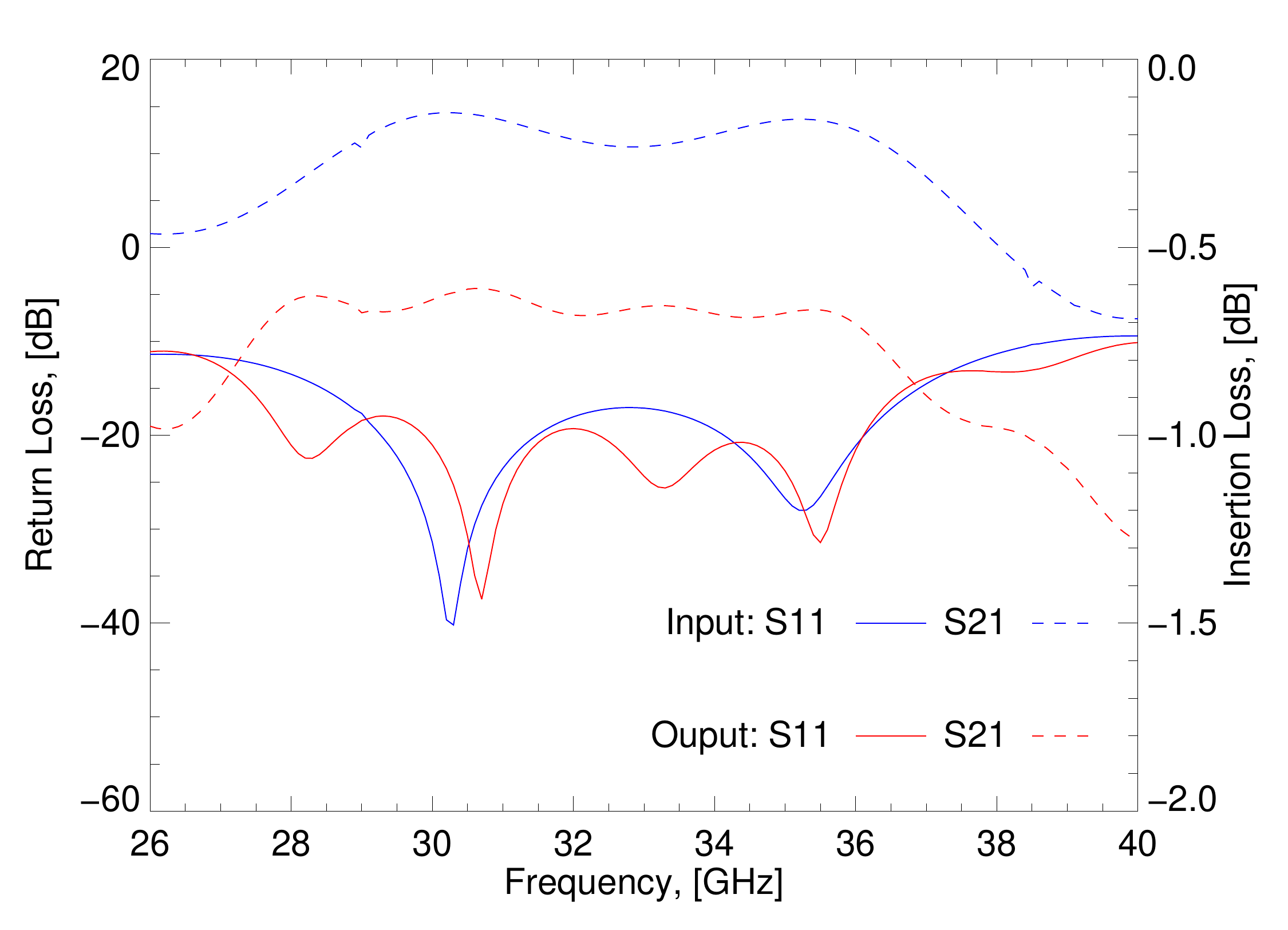}
 \caption{The simulated results for the input and output broadband waveguide to microstrip transition and subsequent 50\,$\Omega$ microstrip lines.}
 \label{fig:probe_performance}
 \end{figure}

To allow the MMIC part of the amplifier behavior to be simulated in both the room temperature and cryogenic model, an equivalent MMIC from the Faraday Project was integrated into a suitable test module. This LNA's S Parameters were measured using an Agilent Technologies PNA-X from which the MMIC's S-parameters were extracted using the de-embedding technique outlined in \cite{agilent}. Since the de-embedding process (\ref{eqn:t_mmic}) requires the S and T  (\ref{eqn:s_to_t_parameters})  parameters of the surrounding test fixture to be known, models of the test module's input and output waveguide-to-microstrip transitions were produced in HFSS (Fig. \ref{fig:probe}) and the S-parameters simulated (Fig. \ref{fig:probe_performance}). The actual de-embedding was performed using the dedicated de-embedding components in ADS.

\begin{equation}
	\begin{bmatrix}
	T_I
	\end{bmatrix}^{-1}
	\begin{bmatrix}T_I
	\end{bmatrix}
	\begin{bmatrix}T_{MMIC}
	\end{bmatrix}
	\begin{bmatrix}T_O
	\end{bmatrix}
	\begin{bmatrix}T_O
	\end{bmatrix}^{-1} = 
	\begin{bmatrix}T_{MMIC}
	\end{bmatrix}
	\label{eqn:t_mmic}
\end{equation}

\begin{equation}
	\begin{bmatrix}
	T_{11} & T_{12} \\
	T{21} & T_{22}
	\end{bmatrix} =
	\frac{1}{S_{21}}
	\begin{bmatrix}
	S_{12}S_{21} - S_{11}S_{22} & S_{11} \\
	-S_{22} & 1
	\end{bmatrix}
\label{eqn:s_to_t_parameters}
\end{equation}

\subsection{Modeling Performance}

\subsubsection{Gain}

The predicted room-temperature and 8-K $S_{21}$ values were compared with the actual measured $S_{21}$ values and can be seen in Fig. \ref{fig:290K_model} and Fig. \ref{fig:20K_model}. The gain measured by the NFM is also shown. The slight ripple in the modeled $S_{21}$ is due to the effects of the long input and output waveguides not being fully removed from the measurements of the MMIC.

\subsubsection{Noise}

\begin{figure}[!h]
 \centering
 \includegraphics[width=3.5in]{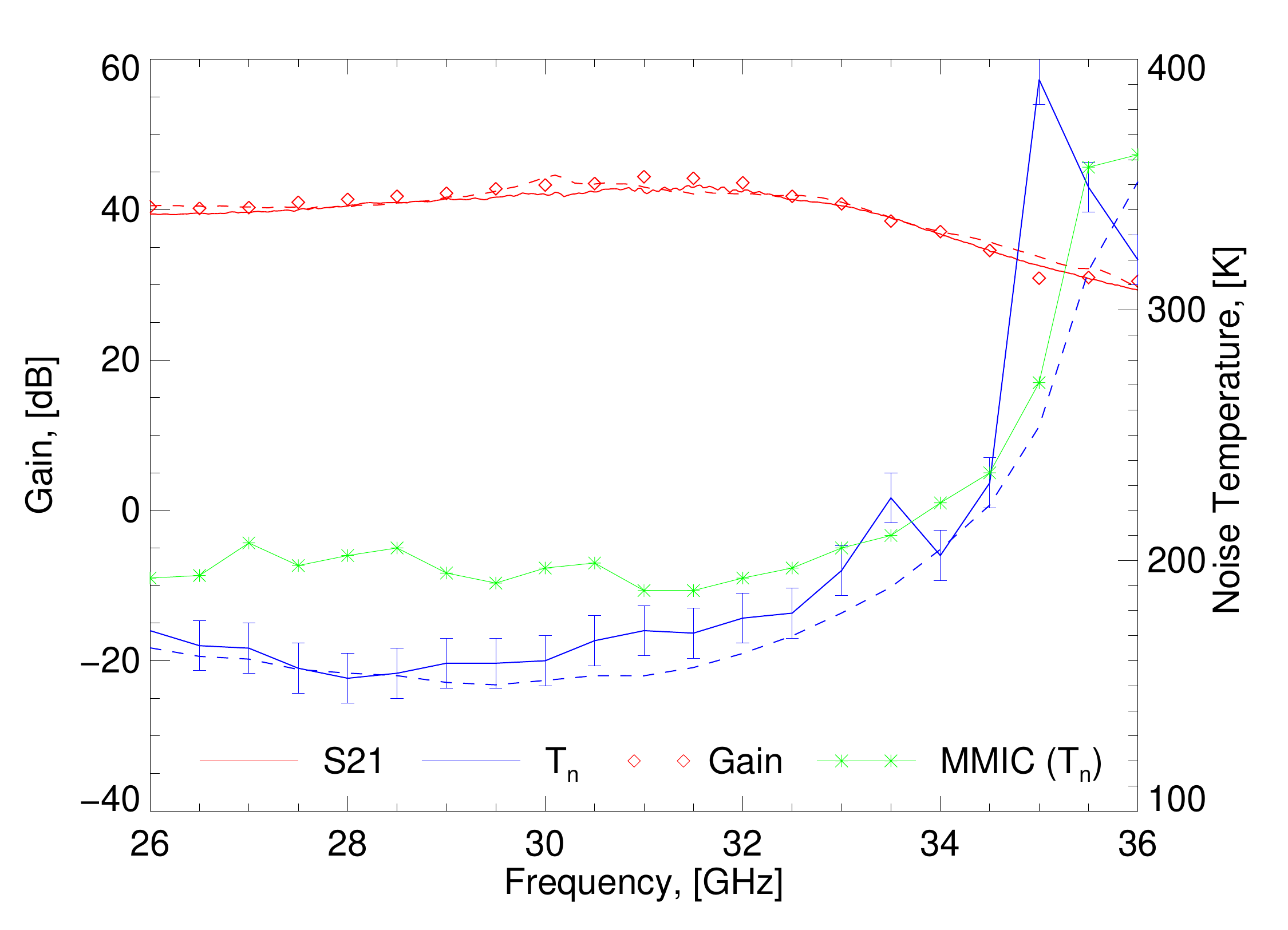}
 \caption{Modeled (dashed line) and measured (solid line) 290-K T+MMIC $S_{21}$ and noise temperature ($T_n$) results for 26--36\,GHz. The T+MMIC's gain recorded by the NFM is also shown, along with the noise performance of the MMIC only amplifier.}
 \label{fig:290K_model}
 \end{figure}

\begin{figure}[!h]
 \centering
 \includegraphics[width=3.5in]{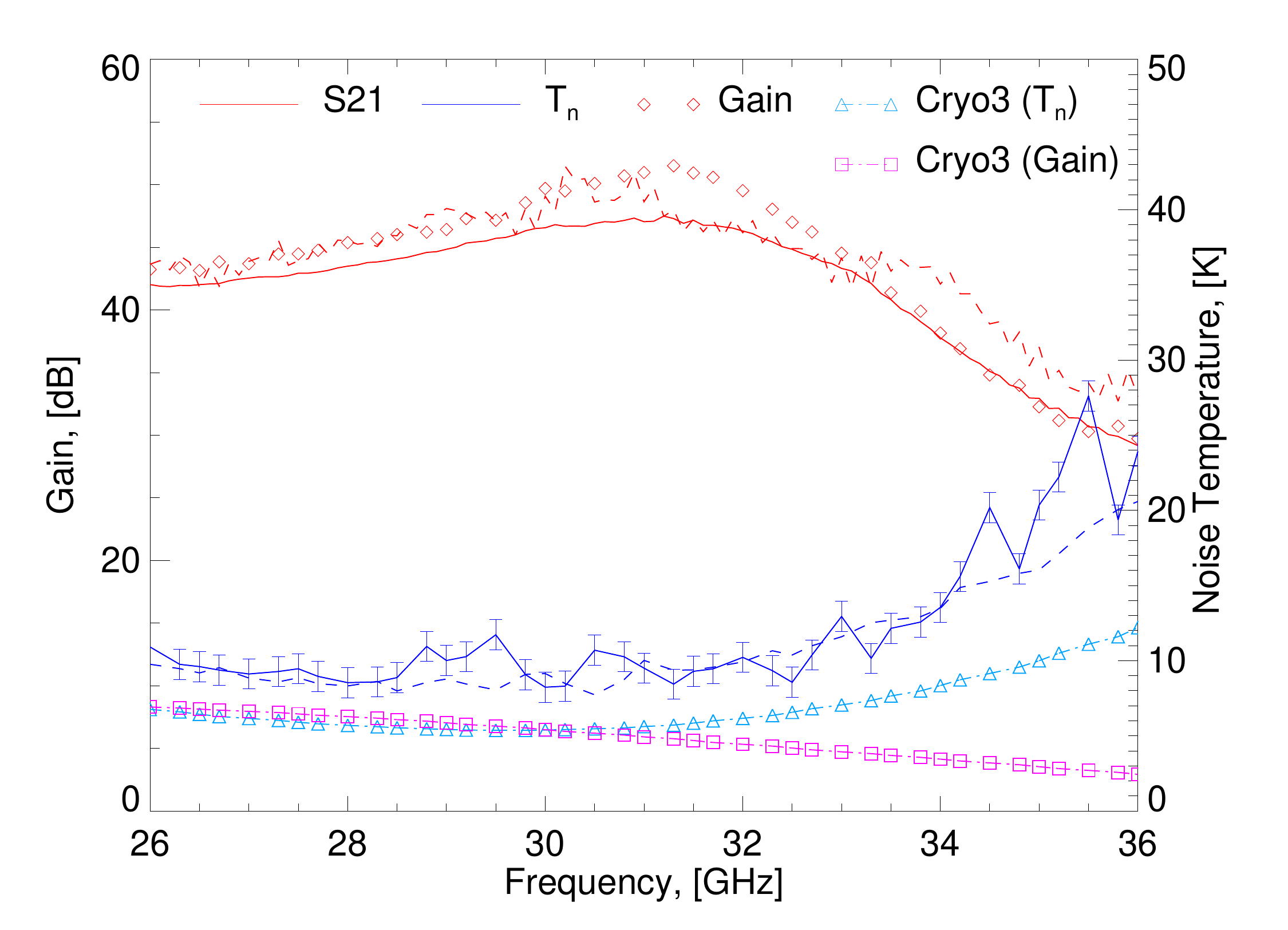}
 \caption{Modeled (dashed line) and measured (solid line) 8-K T+MMIC $S_{21}$ and noise temperature ($T_n$) results for 26--36\,GHz. The modeled noise temperature and the gain of the transistor are also shown, along with the gain of the T+MMIC LNA as recorded by the NFM.}
 \label{fig:20K_model}
 \end{figure}

Using the Pospieszalski noise equivalent temperatures  and the equivalent circuit model, the noise performance of the transistor stage can be modeled. Using this with (\ref{eqn_cascaded_noise}) and the noise performance of the MMIC, the noise performance of the T+MMIC-based LNA can be predicted. The modeled and actual performance are compared in Figs. \ref{fig:290K_model} (290\,K) and \ref{fig:20K_model} (20\,K).

The models confirm that the amplifier is behaving as expected from the Friss equation (\ref{eqn_cascaded_noise}). Fig. \ref{fig:20K_model} shows that within the intended operating bandwidth (27-33\,GHz) the noise temperature of the T+MMIC LNA sits just above the noise temperature of the transistor (first stage), with the gain suppressing the noise contribution of the MMIC. Outside this band, however, once the gain provided by the first stage reduces, the noise of the MMIC becomes more significant and the noise temperature of the T+MMIC LNA drifts towards that of the MMIC, as can be seen in Fig. \ref{fig:290K_model}.


%


\section{Conclusion}

MMIC LNAs are now the preferred choice for the LNAs required by radio astronomy, but their noise performance is still inferior to that of MIC-based LNAs. One possible solution is to use a discrete transistor in front of the MMIC. This paper has reported on the development of such an LNA, with an average noise temperature of 9.4\,K. This is some 4--5\,K lower than an equivalent MMIC LNA, representing a near $50\%$ improvement. Cryogenic cooling to 8\,K has also resulted in an amplifier that almost matches the noise performance of the lowest-noise Ka-band LNAs so far developed, illustrating that cooling below the typical 15--20\,K that is currently used by most radio observatories may prove beneficial. 
We have also presented a simple approach to modeling such an amplifier, showing that the MMIC can almost be regarded as a ``black box" in terms of the amplifier's development with only the transistor's equivalent circuit parameters and noise parameters needing to be measured with a probe station. The modeled data also show that we have demonstrated effective suppression of the (higher) MMIC noise by the lower-noise first-stage transistor, within its operating band.

\section{Discussion}
\begin{figure}[!h]
 \centering
 \includegraphics[trim = 0mm 3cm 0mm 5cm, clip, width=3.5in]{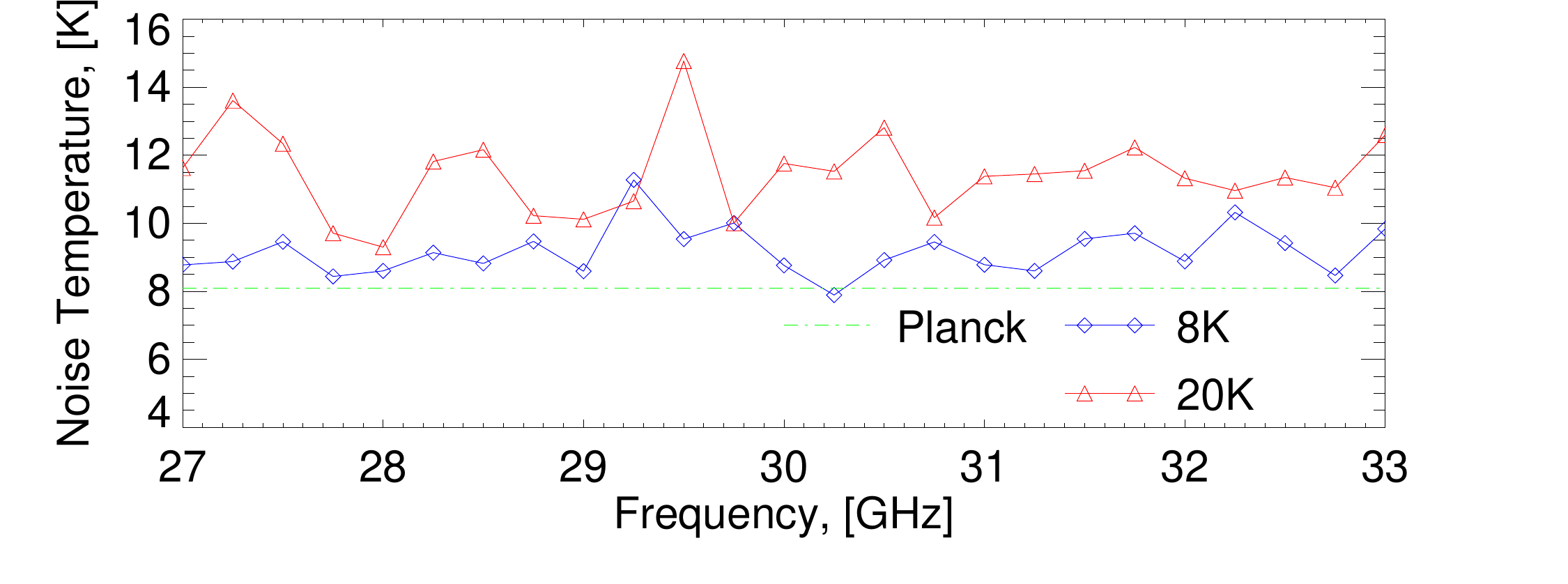}
 \caption{The noise temperature of the T+MMIC LNA at 20\,K and 8\,K, compared to the average noise temperature of the Planck amplifiers at 20\,K.}
 \label{fig:planck}
 \end{figure}

Considering our earlier estimate of the expected noise temperature of the T+MMIC LNA; at 20-K (Fig. \ref{fig:planck}) the noise temperature is slightly higher (11.4\,K) than that expected from the {\it Planck} amplifier (10\,K). This is likely due to the {\it Planck} amplifiers using a slightly different version of the Cryo-3 which had a thiner passivation layer. These transistors were found to have a slightly better noise performance than the type of Cryo-3 transistor that was integrated into this amplifier.

During the development, the LNA was found to oscillate under certain conditions. Examining the measured and modeled S parameters showed that the amplifier was conditionally stable at several fequencies within its design bandwidth. This is likely due to the large quantity of gain and could potentially be resolved by a lower gain MMIC.

This approach would also be applicable to the experiments such as the Q/U-Imaging-Experiment (QUIET) which used integrated modules to observe the CMB, but encountered compression issues when using multiple MMICs in cascade to achieve sufficient gain \cite{reeves2012}. This approach would offer slightly reduced additional gain, thus preventing compression, whilst also offering the potential for reducing the noise temperature.

One obvious drawback of this technology is the need to develop a new module for the integration of the MMIC and the transistor. A preferred approach would be to mount the transistor into its own module and connect via waveguide to an existing MMIC-based amplifier module. 

\section*{Acknowledgment}

The authors would like to thank D. Shepard for machining the LNA module, E. Blackhurst for assembling the LNA and A. Galtress for his help in designing the module. We would also like to thank, Prof P. Wilkinson and Dr D. George for supplying the Faraday MMIC LNA, Prof R Davis for supplying the Cryo-3 transistor, Agilent for supplying ADS and our test equipment and Ansys for supplying HFSS. This work was funded by the Science and Technology Facilities Council (STFC), Consolidated Grant ST/J001562/1.

\ifCLASSOPTIONcaptionsoff
  \newpage
\fi






\bibliographystyle{IEEEtran}
\bibliography{IEEEabrv,Hybrid_LNA_bibliography}

%

%
%

%

\vfill
\eject
\begin{IEEEbiography}[{\includegraphics[width=1in,height=1.25in,clip,keepaspectratio]{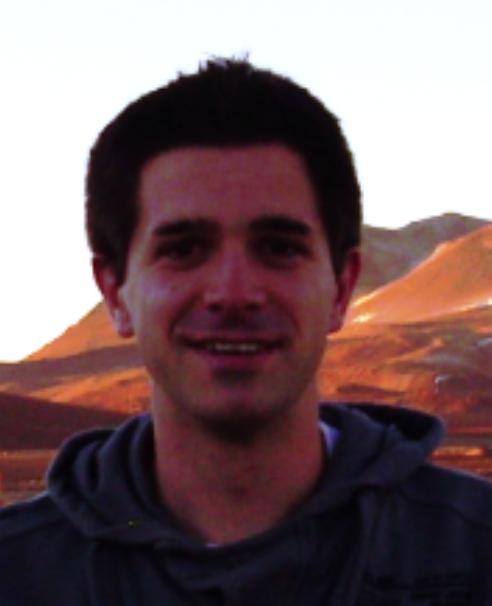}}]{Mark McCulloch}
received his MPhys Physics with Astrophysics degree from the University of Manchester in 2008, and has recently submitted his Ph.D degree thesis at the University of Manchester. For his Ph.D he investigated potential enhancements to LNAs with the aim of lowering the noise temperature of LNAs with special focus to future Cosmic Microwave Background observatories.
\end{IEEEbiography}

\begin{IEEEbiography}[{\includegraphics[width=1in,height=1.25in,clip,keepaspectratio]{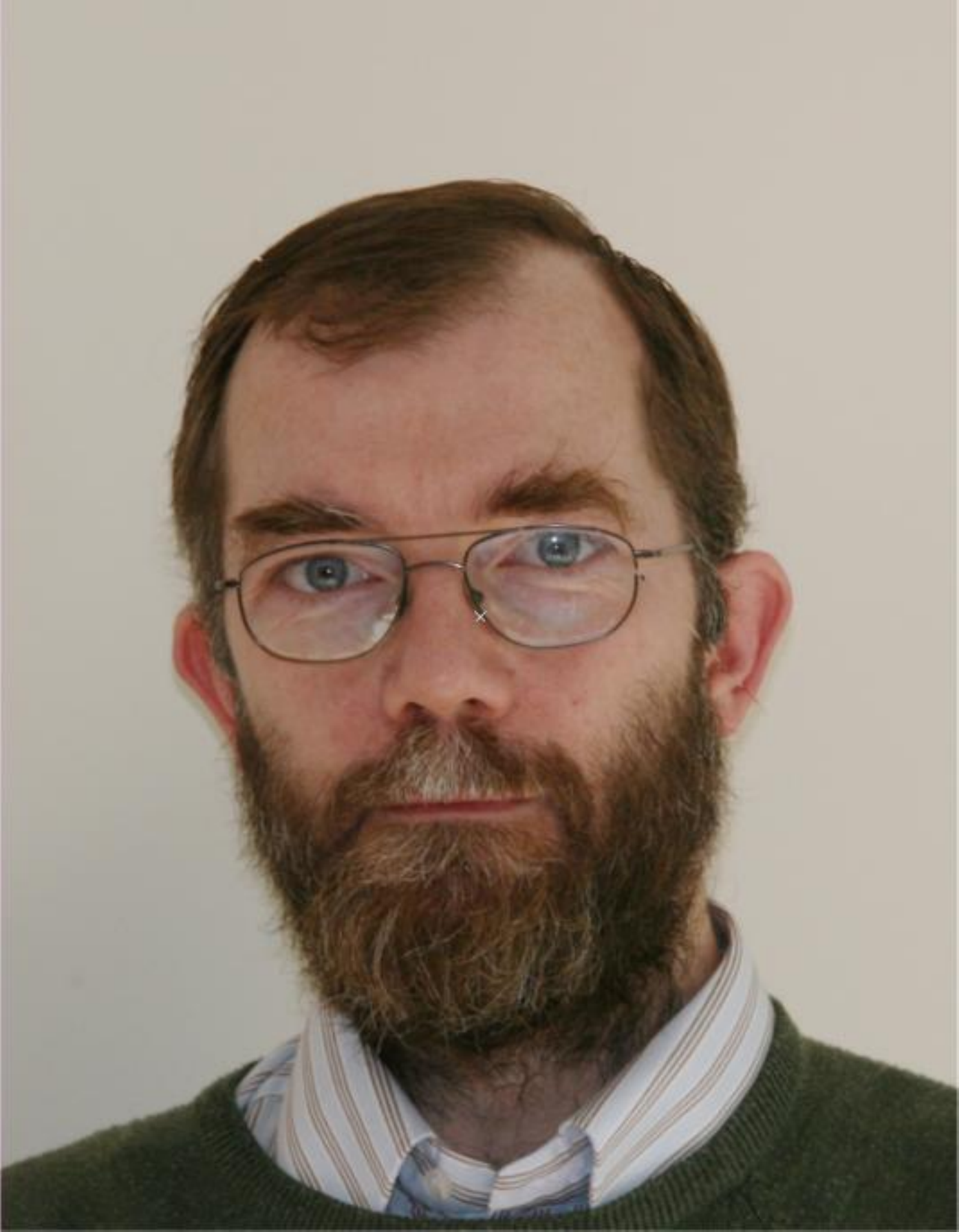}}]{Simon Melhuish}
graduated in Physics from New College, Oxford, and worked
on telescope systems for
Cosmic Microwave Background studies for his Ph.D at Jodrell Bank,
University of Manchester.
He has been responsible for elements of various radio telescope
projects, including the
Very Small Array, QUaD and Clover.
\end{IEEEbiography}

\begin{IEEEbiography}[{\includegraphics[width=1in,height=1.25in,clip, trim = 0 0 0 1cm, keepaspectratio]{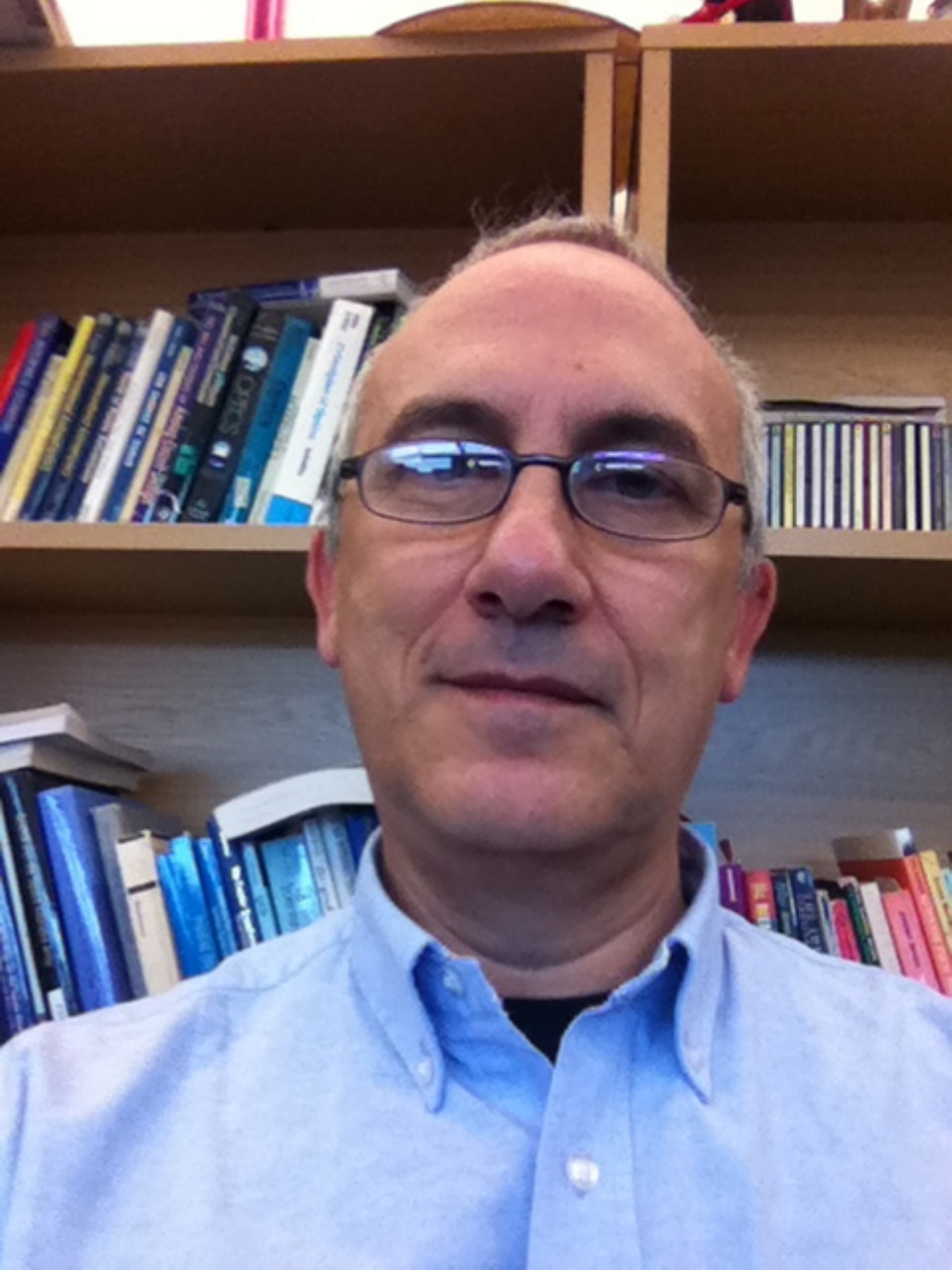}}]{Lucio Piccirillo}
is Professor of Radio Astronomy Technology at the University of Manchester.
His research interests span from experimental cosmology to sub-K cryogenics and cryogenic
Low Noise Amplifiers.
\end{IEEEbiography}




\vfill


\end{document}